\documentclass{article}
\usepackage{spconf,amsmath,graphicx}

\usepackage{hyperref}

\title{LaughNet: synthesizing laughter utterances from waveform silhouettes and a single laughter example}
\name{Hieu-Thi Luong, Junichi Yamagishi\thanks{This study is supported by JST CREST grant JPMJCR18A6 and by MEXT KAKENHI grant 21K19808.}}
\address{National Institute of Informatics, Japan}

\begin{document}
%
\maketitle
\begin{abstract}

Emotional and controllable speech synthesis is a topic that has received much attention. However, most studies focused on improving the expressiveness and controllability in the context of linguistic content, even though natural verbal human communication is inseparable from spontaneous non-speech expressions such as laughter, crying, or grunting. 
We propose a model called LaughNet for synthesizing laughter by using waveform silhouettes as inputs. The motivation is not simply synthesizing new laughter utterances, but testing a novel synthesis-control paradigm that uses an abstract representation of the waveform.
We conducted basic listening test experiments, and the results showed that LaughNet can synthesize laughter utterances with moderate quality and retain the characteristics of the training example. More importantly, the generated waveforms have shapes similar to the input silhouettes. For future work, we will test the same method on other types of human nonverbal expressions and integrate it into more elaborated synthesis systems.

\end{abstract}
\section{Introduction}
\label{sec:intro}

Current speech-synthesis technology can synthesize speech with high naturalness and in a convenient manner for well-contained scenarios such as conversational speech or audiobook narratives.
To push the limitation of the technology further, many studies focused on balancing between developing a seamless end-to-end (E2E) system \cite{wang2017tacotron} and putting back some form of control over synthetic speech \cite{luong2017adapting}. For example, Wang et\ al. \cite{wang2017uncovering} proposed a prosody controlling scheme through manipulating style tokens self-discovered by the sequence-to-sequence text-to-speech (TTS) model, while Shechtman et\ al. \cite{shechtman2021supervised} introduced an emphasis indicator that allows controlling of the prosody of the generated speech.

While speech is essential to human-machine interaction as it allows the exchange of information, non-speech vocalizations, such as laughter, play an important role in expressing emotions \cite{schroder2003experimental}.
Many laughter-synthesis systems use the same methodology as TTS. Specifically, laughter is treated as an extension of speech, thus presented by the same symbolic phonemes \cite{mansouri2019dnn,tits2020laughter} with additional specified contexts \cite{nagata2018defining}. This setup is convenient but inadequate, as a linguistic interface cannot fully express the dynamic of laughter.
Different approaches include training auto-encoder (AE) or variational autoencoder (VAE) \cite{mansouri2020laughter} models that generate laughter from a latent vector, or assuming non-speech vocalizations are realizations of emotions \cite{matsumoto2020controlling}.
Most of these approaches lack the ability to control due to the highly abstract input.

The text in TTS can be seen as the control interface of a synthesis model. In other words, the model learns a mapping between a symbolic input, such as word, character, or phoneme, and a feature-dense output, which can be waveform or acoustic features. By changing the human-comprehensible input, we want to change the generated output.
In this paper, we propose a laughter synthesis model called LaughNet and conducted experiments to test the feasibility of using waveform silhouettes as the interface for controlling.
The rest of the paper is organized as follows: Section \ref{sec:laughnet} describes LaughNet and the motivation behind this research. Section \ref{sec:experiment} provides details about the experimental setup and results. Section \ref{sec:conclusion} concludes the paper and suggests plans for future work. 

\section{LaughNet}
\label{sec:laughnet}

\begin{figure}[t!]
  \centering
  \includegraphics[width=0.75\linewidth]{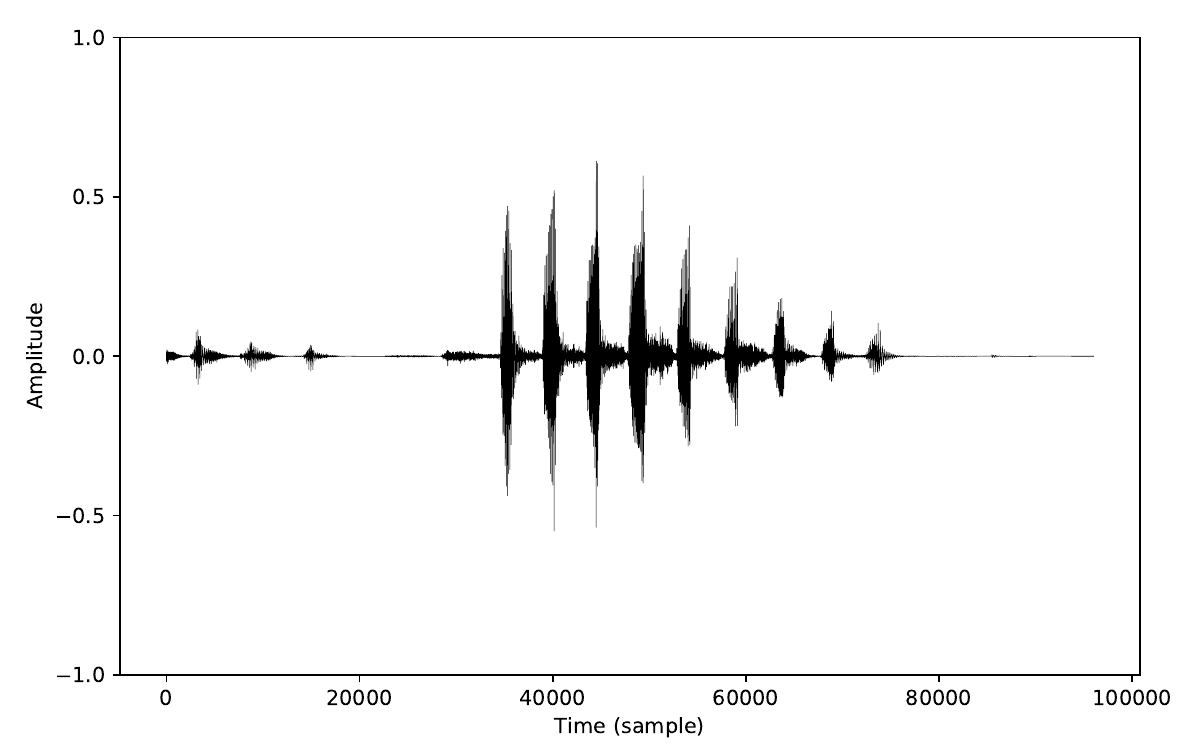}\\
  \vspace{-1mm}
  a) Raw waveform\\
  \vspace{1mm}

  \includegraphics[width=0.75\linewidth]{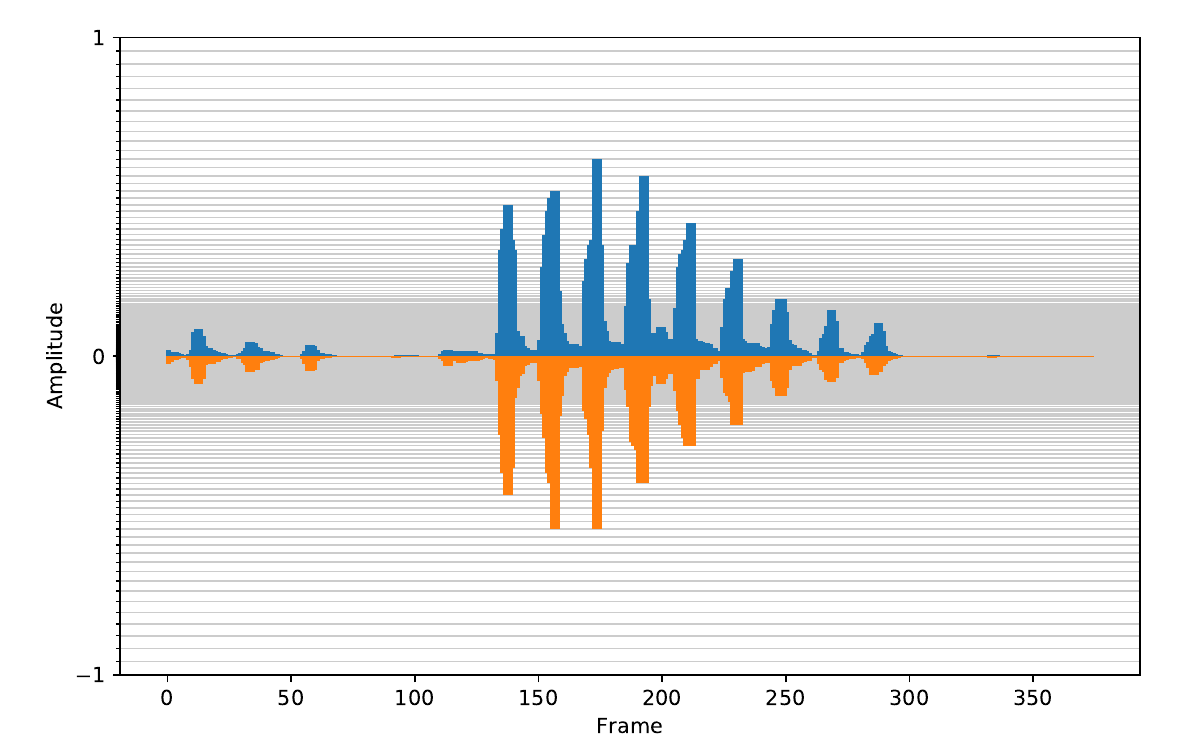}\\
  \vspace{-1mm}
  b) Silhouette, 8-bit $\mu$-law \\
  \vspace{1mm}

  \includegraphics[width=0.75\linewidth]{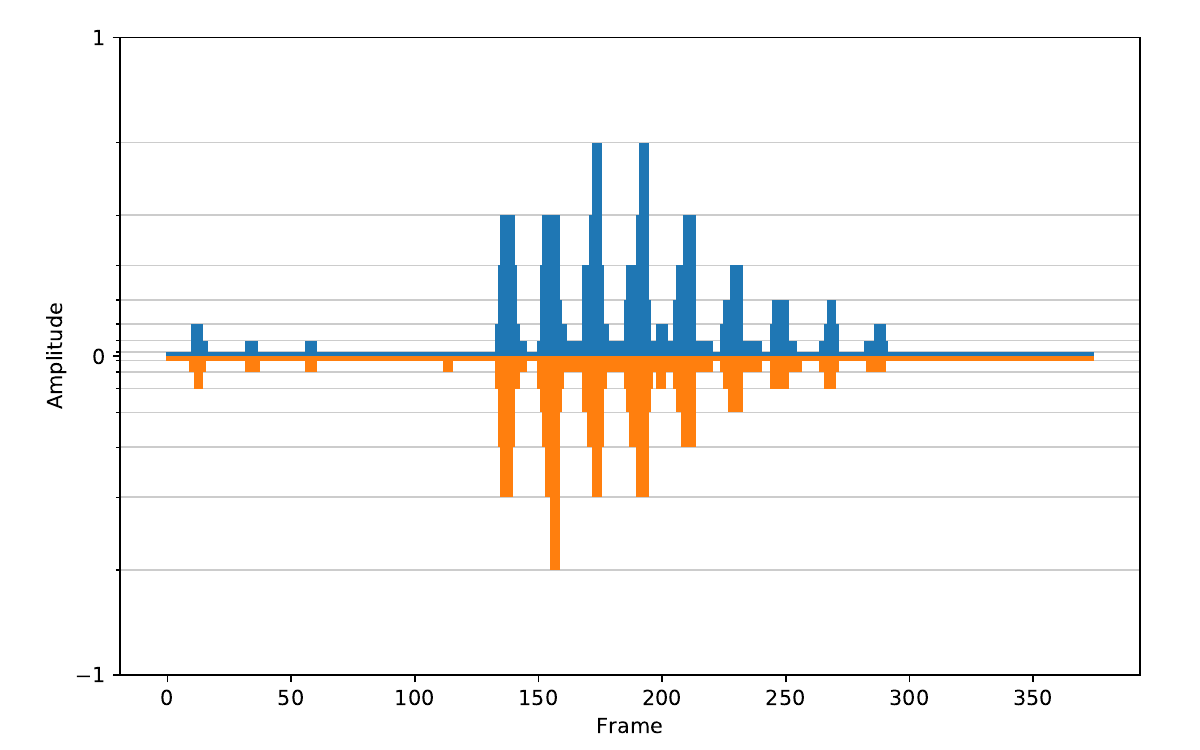}\\
  \vspace{-1mm}
  c) Silhouette, 4-bit $\mu$-law \\
    
  \caption{Waveform silhouettes. Depending on configurations of pooling window and the quantization, silhouette shapes become more or less abstract.}
  \label{fig:waveform_silh}
  \vspace{-3mm}
\end{figure}

\subsection{Waveform silhouettes as synthesis-control interface}

Figure \ref{fig:waveform_silh} illustrates the waveform silhouette used by LaughNet.
Waveform silhouettes are two-dimensional frame-based features extracted from a raw waveform using a series of overlapping windows. This is similar to how a spectrogram is extracted but instead of short-time Fourier transform (STFT), simple max and min pooling operations are used to obtain the `shape' of the waveform as we want an abstract representation. The continuous min and max values are quantized into several bins to further abstract the representation.
By changing the configurations of the pooling window and quantization, we can adjust the level of abstraction of the waveform silhouettes. 
A high-level abstract interface enables more intuitive control for humans but is less detail oriented, and vice versa for a low-level abstract interface.

Investigating different levels of abstraction is the main objective of this research, but as a preliminary study, two levels of quantization, as shown in Figure \ref{fig:waveform_silh}, are tested as the input features for laughter synthesis.
Our study shares a similar motivation with that of Greshler\ et\ al. \cite{greshler2021catch} but instead of changing a low sampling rate waveform to change the generated content, we use an engineered waveform silhouettes as the control interface.

\begin{figure}[t!]
  \centering
  \includegraphics[width=0.85\linewidth]{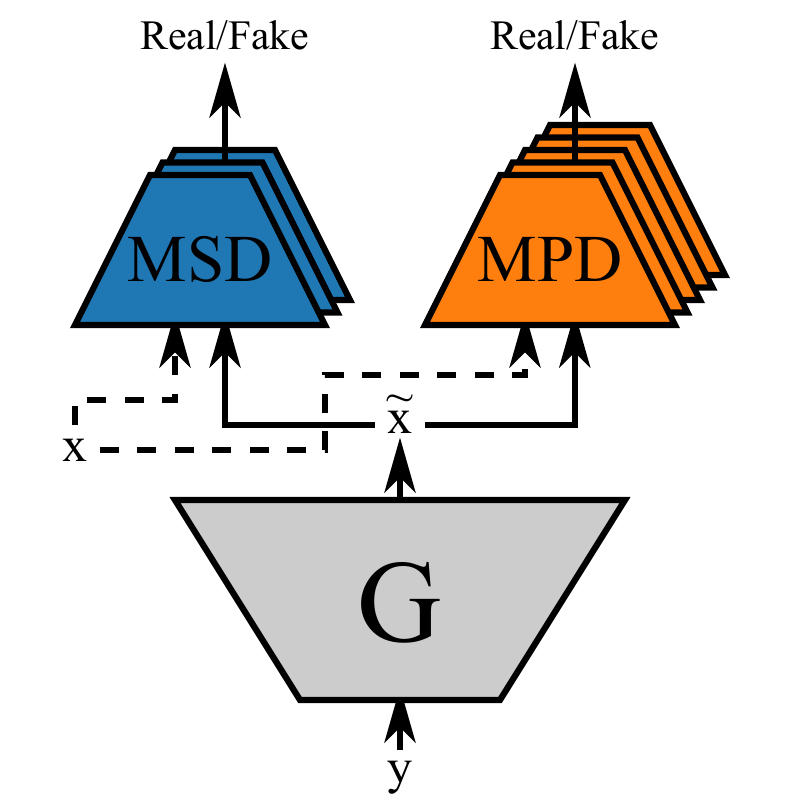}\\
  \vspace{-2mm}
  \caption{LaughNet:  $y$ represents silhouette input, $x$ is real natural waveform, while $\tilde{x}$ is synthetic waveform. Module G is generator, MSD is multiscale discriminator, and MPD is multi-period discriminator. Setup is similar to HiFi-GAN \cite{kong2020hifi} but we use waveform silhouettes instead of mel-spectrogram.}
  \label{fig:laughnet}
  \vspace{-4mm}
\end{figure}

\subsection{Model architecture}

To test our hypothesis, we used HiFi-GAN \cite{kong2020hifi}, a generative adversarial network (GAN)-based \cite{goodfellow2014generative} vocoder \cite{kumar2019melgan} that transforms a mel-spectrogram into a raw waveform, as the basis of our laughter synthesis system. As HiFi-GAN is not autoregressive, it can generate waveforms efficiently and with high-fidelity, which enables us to quickly test our ideas.
The main difference is that we use waveform silhouettes as the input of the generator instead of mel-spectrograms. Readers can refer to the original paper \cite{kong2020hifi} for details. We only summarize the main components in this paper.

The general structure of LaughNet is illustrated in Figure \ref{fig:laughnet}. The generator (G) takes a frame-based silhouette and up-samples it to the designated sampling rate. G consists of many up-sampling layers and several multi-receptive field fusion modules that add features from residual blocks of different kernel sizes and dilation rates. The amount of parameters can be adjusted, but for LaughNet model, we simply use the $V1$ configuration described in the original study \cite{kong2020hifi}.
In a typical GAN-based fashion, G was trained alternatively with one or several discriminators in a minimax game. The same as HiFi-GAN, our model includes two types of discriminators: the multi-period discriminator which learns to discriminate real and fake segments by processing samples at a particular period; and the multiscale discriminator \cite{binkowski2019high,kumar2019melgan} which learns to discriminate by processing samples at a down-sampling scale.
To train G, both feature-matching loss \cite{kumar2019melgan} and mel-spectrogram loss \cite{kong2020hifi} are used.

\section{Experiments}
\label{sec:experiment}

\begin{figure}[t!]
  \centering
  \includegraphics[width=0.85\linewidth]{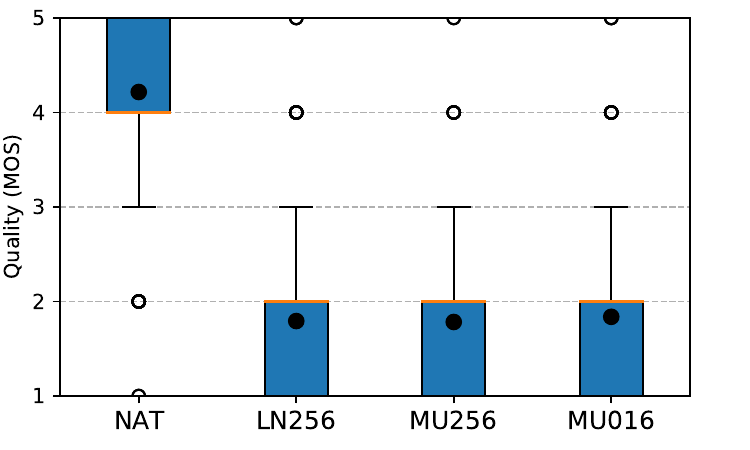}\\
  \vspace{-2mm}
  a) Quality (5-point scale)\\
  \vspace{1mm}

  \centering
  \includegraphics[width=0.85\linewidth]{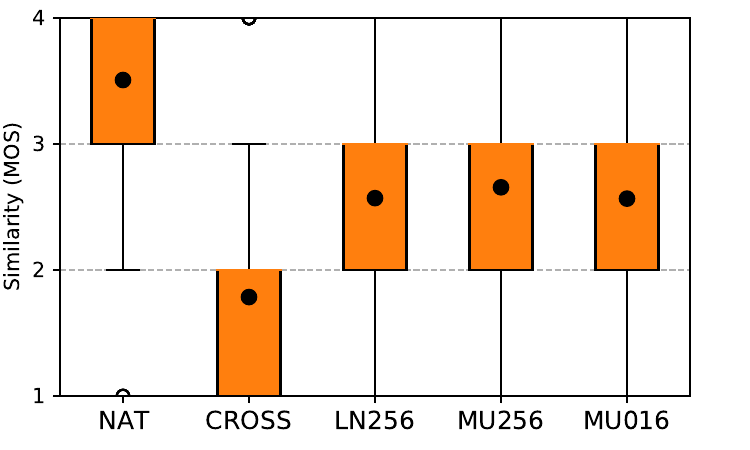}\\
  \vspace{-2mm}
  a) Similarity (4-point scale)\\
  \caption{Subjective evaluations. NAT is the natural laughter sample of the target. CROSS is a sample of a speaker with the opposite sex, and \texttt{LN256}, \texttt{MU256}, and \texttt{MU016} are slightly different versions of LaughNet.}
  \label{fig:subjective}
  \vspace{-3mm}

\end{figure}

\subsection{Laughter data}
As high-quality conversational laughter data \cite{mori2019conversational} is spontaneous and difficult to obtain, we decided to used acted laughter, which have clear emotion intents and more isolated, for our experiments.
Specifically, we used video-game-development assets purchased from Unity Asset Store\footnote{\url{https://assetstore.unity.com/}}. It is a common practice for small and independent game developers to use these assets in their games.
Video game development is also a practical application scenario of the proposed system.
The Laugh SFX package we purchased has laughter of several male and female actors. For training, eight utterances, four from male and four from female, were used as the targets.
As these sound assets are not designed for research, information about speakers were not provided, we simply selected utterances with perceivable differences in voice characteristics as different targets.
Another set of eight laughter utterances were selected as template, or `source' utterances, the silhouettes of which were used as the inputs for synthesizing laughter utterances.
Our laughter synthesis system, in a way, can be regarded as a very special voice conversion (VC) system \cite{sisman2020overview}, hence the use of similar terminologies. The experiment and the evaluation were also structured similar to that of VC.

As we need to train a laughter synthesis system with a single example, to achieve a stable and decent performance, we first train the model with a multi-speaker speech corpus. Specifically, we used 24.4 hours of speech from 100 speakers of the VCTK corpus \cite{veaux2017superseded}. Prior studies showed that transfer learning from speech can help improve the performance of a laughter-synthesis model \cite{tits2020laughter}.

\subsection{Training and evaluating setups}

We used 24-kHz waveforms for the experiments. A sliding window with 1024 samples in length and moving 256 points each step was used to extract silhouettes. Each frame is a two-dimensional feature vector consists of the min and max value within the window.
The continuous values were then quantized into a set of values. We split them into equally distributed bins (linear) or using the $\mu$-law algorithm. Three versions of LaughNet were evaluated: linear 256-bin \texttt{LN256}, 8-bit $\mu$-law \texttt{MU256}, and 4-bit $\mu$-law \texttt{MU016}.
The hyperparameters used in training were the same as those used for HiFi-GAN \cite{kong2020hifi} which includes both a mel-spectrogram and a feature-matching losses in the generator optimization loss. Mel-spectrograms were extracted using the same sliding-window configuration as the waveform silhouette.

We first initialized these using VCTK speech data for 150000 steps with each training batch consisting of 16 6-second segments randomly extracted from training utterances. Waveforms were also randomly scaled by a factor $\lambda \in [0.3,1.0]$. This helps increase the variety of waveform silhouettes.
We then fine-tuned the models to the target laughter for another 50000 steps.
For evaluation, we extracted two 6-second segments from each source utterance at a random position to create a test set of 16 silhouettes.
Even though the end goal is creating a system in which users can control the synthetic laughter by directly changing the input, we used silhouettes extracted from source laughter utterances for testing, as this was a preliminary study.
Given this arrangement, our experiment was quite similar to voice conversion, as we essentially wanted to change the characteristics of the laughter while retaining the `shape' of the source waveform. In summary, each version of LaughNet created 8 models by fine-tuning to the 8 target laughter utterances, which were used to generate new laughter utterances from 16 silhouettes extracted from source laughter utterances.

\subsection{Quality and speaker similarity evaluations}

\begin{table}[t]
    \caption{MSEs between the input silhouettes and silhouettes of the synthesized laughter waveforms.}
    \label{tab:mse}

    \centering
    \vspace{3mm}
    \scalebox{1.0}{
    \begin{tabular}{llrrr}
        \hline
        & Type & nBins & nTests & MSE \\ \hline
        \texttt{LN256} & linear & 256 & 128 & 0.0354 \\ \hline
        \texttt{MU256} & $\mu$-law & 256 & 128 & 0.0336 \\ \hline
        \texttt{MU016} & $\mu$-law & 16 & 128 & 0.0336 \\ \hline
    \end{tabular}}\\
    \vspace{-2mm}
\end{table}

We evaluated three versions of LaughNet by conducting listening tests in which participants were asked to judge the quality and speaker similarity of the laughter samples. We gathered 16 participants, each did 8 sessions, in which 16 quality and 16 similarity questions were included.
For the quality evaluation, listeners were presented with a natural laughter sample (NAT) of a target or a sample from one of the synthesis models and asked to judge quality. Figure \ref{fig:subjective}a shows the results in which all of the models received low score with \texttt{MU016} scoring slightly better than the other two.
For the similarity evaluation, listeners were presented with two samples, one from a target and the other from one of the evaluated models. These included natural samples of the same target (NAT) and the natural laughter sample of a random speaker with a different sex than the target (CROSS).
Figure \ref{fig:subjective}b shows the results of the similarity evaluation, in which all three LaughNet models had lower scores than NAT but higher than CROSS. Out of all three LaughNet models, \texttt{MU256} scored slightly better than the other two.

In summary, the synthesized laughter samples had low quality, but express a certain level of individuality as indicated from the similarity-evaluation results.

\subsection{Waveform silhouettes as control interfaces}

Since the purpose of this research was not simply synthesizing new laughter utterances but developing a novel control interface for synthesis systems, we need a method to evaluate the performance of waveform silhouettes as means of control.
Table \ref{tab:mse} shows the mean squared error (MSE) between the input silhouettes and silhouettes of the generated waveforms. More specifically, the error was calculated using the silhouette before quantization. Simply speaking, if we had an accuracy interface, then the error between the input and output silhouette would be minimal. Interestingly, the two models used $\mu$-law for quantization (\texttt{MU256} and \texttt{MU016}) had the same error, which is lower than the model using the linear function for quantization (\texttt{LN256}).
We can say that \texttt{MU256} and \texttt{MU016} have a more precise control interface than \texttt{LN256}. 

\begin{figure}[t!]
  \centering
  \includegraphics[width=0.75\linewidth]{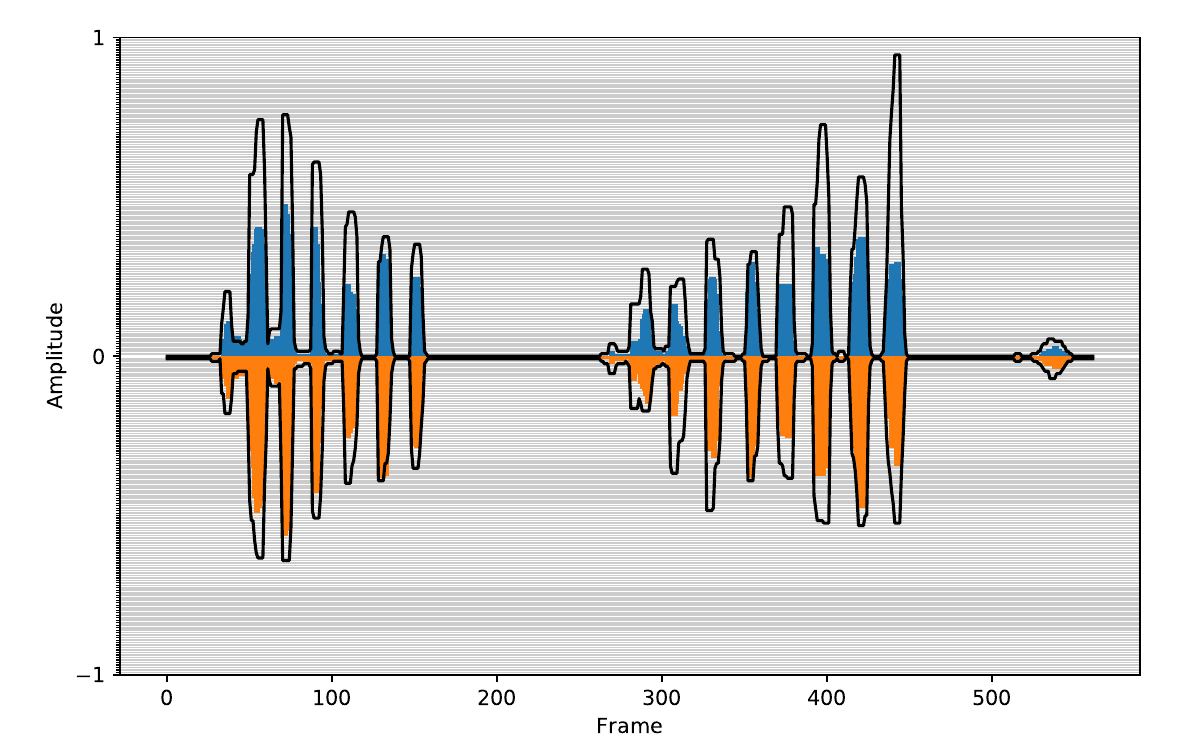}\\
  \vspace{-1mm}
  a) \texttt{LN256}\\
  \vspace{1mm}

  \includegraphics[width=0.75\linewidth]{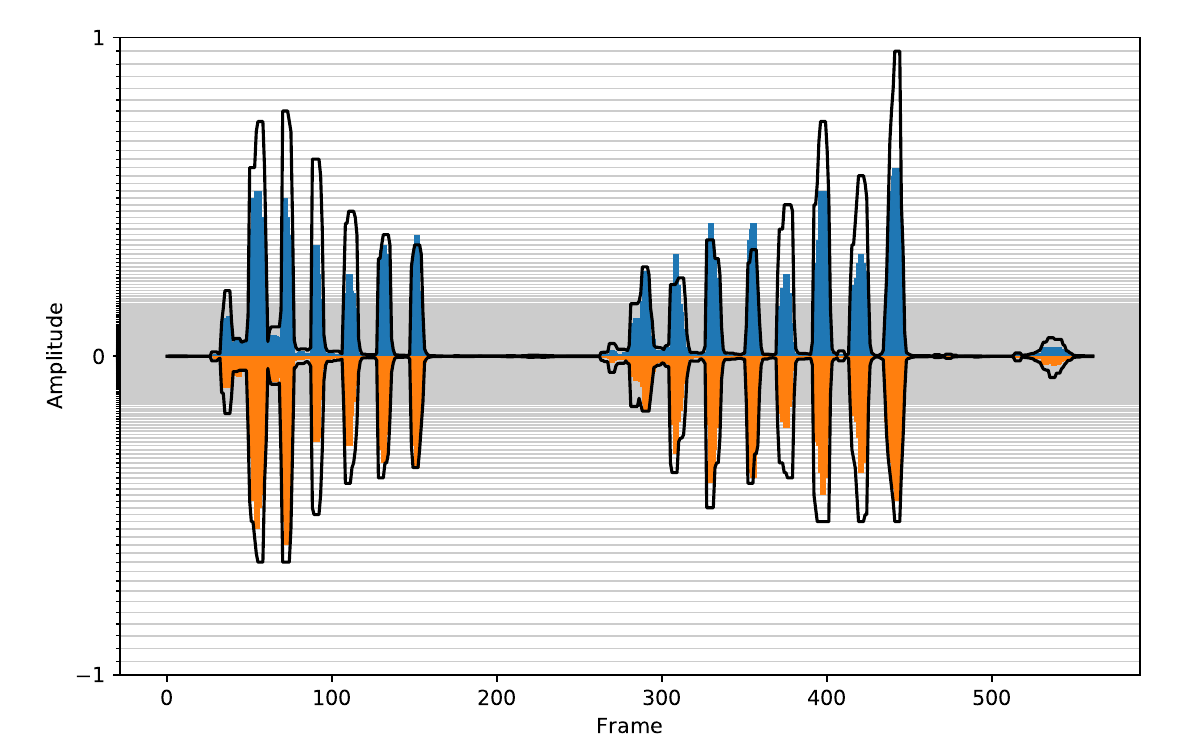}\\
  \vspace{-1mm}
  b) \texttt{MU256} \\
  \vspace{1mm}

  \includegraphics[width=0.75\linewidth]{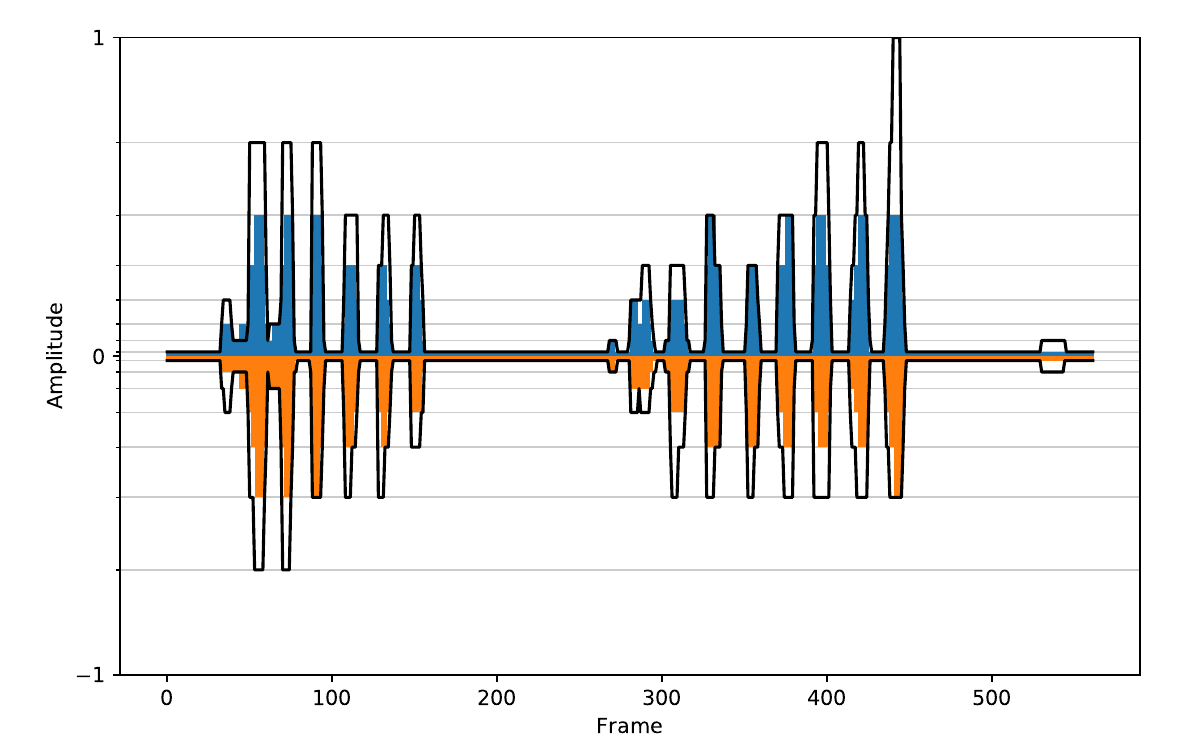}\\
  \vspace{-1mm}
  c) \texttt{MU016} \\
  \vspace{-2mm}
  \caption{Silhouettes of generated laughing utterances. Black borders indicate input silhouettes, while colored bars are the silhouettes of synthetic waveforms.}
  \label{fig:syn_silh}
  \vspace{-3mm}
\end{figure}

While MSE provide some objective evaluations over the control accuracy, it is difficult for humans to understand. Figure \ref{fig:syn_silh} presents examples of input silhouettes and silhouettes of the synthesized laughter. 
Subjectively speaking, we can see that the input silhouette successfully guided the shape of the generated waveform.
Furthermore, we performed simple manipulation operations (stretch the silhouettes vertical and horizontally) to test the feasibility to use waveform silhouette as a mean of control. These samples were not evaluated subjectively, but readers can listen to them in the associated web page\footnote{Laughter samples are available at \url{https://nii-yamagishilab.github.io/sample-laughnet-waveform-silhouette/}}.
It is important to note that the accuracy of the control needs to be put in the context of a synthesis system, meaning that while we want an accuracy control, we also want LaughNet to synthesize laughter with the characteristics of the training example. Balancing between the two objectives is a research topic that we want to explore more in the future. One direction to solve this problem is increasing the level of abstraction of the waveform silhouette and allow the synthesis model to automatically determine the details of the waveform

\section{Conclusion}
\label{sec:conclusion}

We proposed a model for synthesizing laughter utterances using waveform silhouettes as input.
The experimental results indicate that LaughNet can not only synthesize laughter with moderate quality and characteristics of the target speaker with just a single training example, but also have a reliable control interface that can dictate the shape of the generated waveform.
This research is a pioneering regarding exploring novel interfaces for controllable synthesis systems.
For future work, we will extend the same method to other human sounds, such as grunting, spontaneous filler words \cite{szekely2019train}, and affect bursts \cite{schroder2003experimental}, and integrate it into a more complex system \cite{mori2019conversational}. While these non-speech sounds do not convey linguistic information, they do carry a clear and identifiable emotional meaning. It will be inadequate to develop an emotional-speech-synthesis system or natural human-machine interaction but excluding these expressions.

\bibliographystyle{IEEEbib}
\bibliography{refs}

\begin{thebibliography}{10}

\bibitem{wang2017tacotron}
Yuxuan Wang, RJ~Skerry-Ryan, Daisy Stanton, Yonghui Wu, Ron~J Weiss, Navdeep
  Jaitly, Zongheng Yang, Ying Xiao, Zhifeng Chen, Samy Bengio, Quoc Le, Yannis
  Agiomyrgiannakis, Rob Clark, and Rif~A Saurous,
\newblock ``Tacotron: Towards end-to-end speech synthesis,''
\newblock in {\em Proc. INTERSPEECH}, 2017, pp. 4006--4010.

\bibitem{luong2017adapting}
Hieu-Thi Luong, Shinji Takaki, Gustav~Eje Henter, and Junichi Yamagishi,
\newblock ``Adapting and controlling dnn-based speech synthesis using input
  codes,''
\newblock in {\em Proc. ICASSP}, 2017, pp. 4905--4909.

\bibitem{wang2017uncovering}
Yuxuan Wang, RJ~Skerry-Ryan, Ying Xiao, Daisy Stanton, Joel Shor, Eric
  Battenberg, Rob Clark, and Rif~A Saurous,
\newblock ``Uncovering latent style factors for expressive speech synthesis,''
\newblock in {\em Proc. ML4Audio Workshop, NIPS}, 2017.

\bibitem{shechtman2021supervised}
Slava Shechtman, Raul Fernandez, and David Haws,
\newblock ``Supervised and unsupervised approaches for controlling narrow
  lexical focus in sequence-to-sequence speech synthesis,''
\newblock in {\em Proc. SLT}, 2021, pp. 431--437.

\bibitem{schroder2003experimental}
Marc Schr{\"o}der,
\newblock ``Experimental study of affect bursts,''
\newblock {\em Speech communication}, vol. 40, no. 1-2, pp. 99--116, 2003.

\bibitem{mansouri2019dnn}
Nadia Mansouri and Zied Lachiri,
\newblock ``Dnn-based laughter synthesis,''
\newblock in {\em Proc. ICCAD}, 2019, pp. 1--6.

\bibitem{tits2020laughter}
Noé Tits, Kevin~El Haddad, and Thierry Dutoit,
\newblock ``{Laughter Synthesis: Combining Seq2seq Modeling with Transfer
  Learning},''
\newblock in {\em Proc. INTERSPEECH}, 2020, pp. 3401--3405.

\bibitem{nagata2018defining}
Tomohiro Nagata and Hiroki Mori,
\newblock ``Defining laughter context for laughter synthesis with spontaneous
  speech corpus,''
\newblock {\em IEEE Transactions on Affective Computing}, vol. 11, no. 3, pp.
  553--559, 2018.

\bibitem{mansouri2020laughter}
Nadia Mansouri and Zied Lachiri,
\newblock ``Laughter synthesis: A comparison between variational autoencoder
  and autoencoder,''
\newblock in {\em Proc. ATSIP}, 2020, pp. 1--6.

\bibitem{matsumoto2020controlling}
Kento Matsumoto, Sunao Hara, and Masanobu Abe,
\newblock ``Controlling the strength of emotions in speech-like emotional sound
  generated by wavenet.,''
\newblock in {\em Proc. INTERSPEECH}, 2020, pp. 3421--3425.

\bibitem{greshler2021catch}
Gal Greshler, Tamar~Rott Shaham, and Tomer Michaeli,
\newblock ``Catch-a-waveform: Learning to generate audio from a single short
  example,''
\newblock {\em arXiv preprint arXiv:2106.06426}, 2021.

\bibitem{kong2020hifi}
Jungil Kong, Jaehyeon Kim, and Jaekyoung Bae,
\newblock ``Hifi-gan: Generative adversarial networks for efficient and high
  fidelity speech synthesis,''
\newblock in {\em Proc. NeurIPS}, 2020, pp. 1--12.

\bibitem{goodfellow2014generative}
Ian Goodfellow, Jean Pouget-Abadie, Mehdi Mirza, Bing Xu, David Warde-Farley,
  Sherjil Ozair, Aaron Courville, and Yoshua Bengio,
\newblock ``Generative adversarial nets,''
\newblock {\em Proc. NIPS}, pp. 1--9, 2014.

\bibitem{kumar2019melgan}
Kundan Kumar, Rithesh Kumar, Thibault de~Boissiere, Lucas Gestin, Wei~Zhen
  Teoh, Jose Sotelo, Alexandre de~Br{\'e}bisson, Yoshua Bengio, and Aaron~C
  Courville,
\newblock ``Melgan: Generative adversarial networks for conditional waveform
  synthesis,''
\newblock in {\em Proc. NeurIPS}, 2019, pp. 1--12.

\bibitem{binkowski2019high}
Miko{\l}aj Bi{\'n}kowski, Jeff Donahue, Sander Dieleman, Aidan Clark, Erich
  Elsen, Norman Casagrande, Luis~C Cobo, and Karen Simonyan,
\newblock ``High fidelity speech synthesis with adversarial networks,''
\newblock in {\em Proc. ICLR}, 2019.

\bibitem{mori2019conversational}
Hiroki Mori, Tomohiro Nagata, and Yoshiko Arimoto,
\newblock ``Conversational and social laughter synthesis with wavenet.,''
\newblock in {\em Proc. INTERSPEECH}, 2019, pp. 520--523.

\bibitem{sisman2020overview}
Berrak Sisman, Junichi Yamagishi, Simon King, and Haizhou Li,
\newblock ``An overview of voice conversion and its challenges: From
  statistical modeling to deep learning,''
\newblock {\em IEEE/ACM Trans. Audio, Speech, Language Process.}, 2020.

\bibitem{veaux2017superseded}
Christophe Veaux, Junichi Yamagishi, and Kirsten MacDonald,
\newblock ``{CSTR VCTK} corpus: English multi-speaker corpus for {CSTR} voice
  cloning toolkit,''
\newblock 2017,
\newblock http://dx.doi.org/10.7488/ds/1994.

\bibitem{szekely2019train}
{\'E}va Sz{\'e}kely, Gustav~Eje Henter, Jonas Beskow, and Joakim Gustafson,
\newblock ``How to train your fillers: uh and um in spontaneous speech
  synthesis,''
\newblock in {\em Proc. SSW}, 2019.

\end{thebibliography}

\end{document}